# A monitoring and diagnostic approach for stochastic textured surfaces

Anh Tuan Bui and Daniel W. Apley

Department of Industrial Engineering and Management Sciences
Northwestern University, Evanston, IL, USA

We develop a supervised-learning-based approach for monitoring and diagnosing texture-related defects in manufactured products characterized by stochastic textured surfaces that satisfy the locality and stationarity properties of Markov random fields. Examples of stochastic textured surface data include images of woven textiles; image or surface metrology data for machined, cast, or formed metal parts; microscopy images of material microstructure samples; etc. To characterize the complex spatial statistical dependencies of in-control samples of the stochastic textured surface, we use rather generic supervised learning methods, which provide an implicit characterization of the joint distribution of the surface texture. We propose two spatial moving statistics, which are computed from residual errors of the fitted supervised learning model, for monitoring and diagnosing local aberrations in the general spatial statistical behavior of newly manufactured stochastic textured surface samples in a statistical process control context. We illustrate the approach using images of textile fabric samples and simulated 2-D stochastic processes, for which the algorithm successfully detects local defects of various natures. Supplemental discussions, results, data and computer codes are available online.

---

Corresponding author: Daniel W. Apley. Email: apley@northwestern.edu.

## 1. Introduction

Image and other profile data are increasingly commonly collected for manufacturing quality control purposes. In this paper, we consider a subcategory of such data that we refer to as stochastic textured surface data, which can be viewed as 2-D stochastic processes. For example, Figure 1(a) is an image of a textile material with enough magnification to show the weave patterns, which exhibit a great deal of stochastic behavior and are not deterministically positioned. Figure 1(b) is a greyscale image version of a simulated 2-D stochastic process sample that could represent surface roughness of a fabricated part. We consider both of these examples later. Other examples of the stochastic textured surface data include images of stone countertops (Liu and MacGregor 2006), ceramic capacitor surfaces (Lin 2007a), lumber surfaces (Bharati, MacGregor, and Tropper 2003), and microscopy images of material microstructure samples (Torquato 2002, Liu and Shapiro 2015). Point cloud surface roughness data of machined, cast, or formed metal parts, obtained from either a contact stylus trace or optical laser scanning, is another example of the stochastic textured surface data. Throughout this paper, we illustrate the approach with image data.

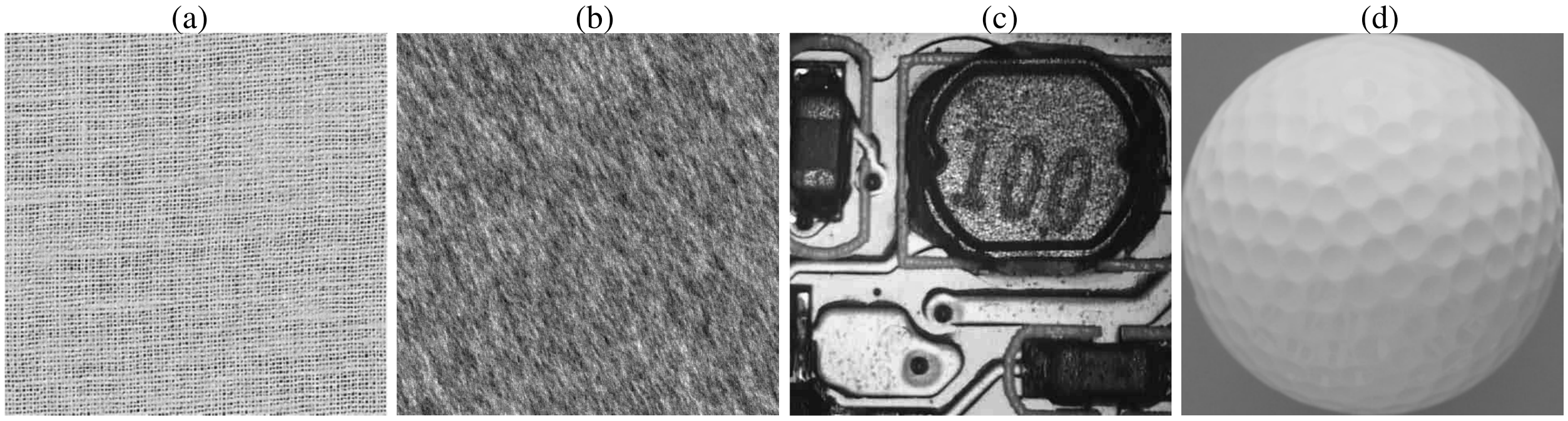

**Figure 1**. Images of (a) a textile fabric sample, (b) a simulated 2-D stochastic process, (c) components on a circuit board, and (d) a golf ball. The first two are examples of stochastic textured surfaces to which the approach of this paper applies.

The stochastic textured surface data have a distinguishing characteristic that renders most of the statistical process control (SPC) literature for profile data largely inapplicable. Such profile

SPC works include Woodall et al. (2004), Zou, Tsung, and Wang (2007), Zou, Wang, and Tsung (2008), Chicken, Pignatiello, and Simpson (2009), Jensen and Birch (2009), Chang and Yadama (2010), Qiu, Zou, and Wang (2010), Qiu and Zou (2010), Wells et al. (2012), Xu et al. (2012), Yu, Zou, and Wang (2012), Zou, Ning, and Tsung (2012), Paynabar, Jin, and Pacella (2013), Viveros-Aguilera, Steiner, and MacKay (2014), Grasso, Colosimo, and Secchi (2016), and Paynabar, Zou, and Qiu (2016). Profile monitoring works such as these are inapplicable to our stochastic textured surface problem, because they require images (i.e., 2-D profiles) for which there is some "gold standard" image comprised of distinct features that represent normal behavior of the process. Typically, the gold standard image would be closely related to the mean image, where the mean is across a sample of multiple images at the same location. For example, in detecting missing or miss-positioned components in a printed circuit assembly, the gold standard is an image of a complete assembly with all the components assembled correctly, as shown in Figure 1(c). In defect defection on smooth metallic surfaces, the gold standard is trivially a non-textured surface of a constant intensity. In monitoring stamping press tonnage signatures (Jin and Shi 1999), the gold standard is the ideal tonnage signature over the course of one stamping cycle that results when the process is behaving normally. For monitoring surfaces that have deterministically repeated patterns under ideal behavior, such as the dimpled surface of the golf ball in Figure 1(d), the gold standard is an image of a golf ball from computer aided design representations or from normal process behavior, perhaps after properly registering and aligning (for image registration techniques, see Xing and Qiu 2011, Qiu and Xing 2013a, Qiu and Xing 2013b).

In stark contrast, there is no such gold standard image for stochastic textured surfaces like those depicted in Figures 1(a) and (b), because the specific configuration of pixel greyscale values varies stochastically from image to image under normal process behavior. In other words, there are infinitely many stochastic textured surface images that have exactly the same normal stochastic behavior, but are all completely different images that do not match pixel-to-pixel. Moreover, they cannot be easily aligned, transformed or warped into a common gold standard image, because of the stochastic nature of the surface. One might consider defining the gold standard image as the spatial mean function for the image (where the mean is taken across an ensemble of images of the same size). However, because of the stationary stochastic nature of the surfaces we consider, the spatial mean function for an image would have the same constant greyscale intensity value for every pixel in the entire image. In other words, the only possible gold standard image would have the same greyscale intensity for every pixel, and any comparison of the inspection images to this gold standard image would have little relevance for detecting defects.

Standard parametric random field models such as Gaussian random fields (Rasmussen and Williams 2006) lack the flexibility to capture the complex dynamics of many stochastic textured surfaces. For example, the warps and wefts of the textile in Figure 1(a) have components that resemble spatial periodicity, but their distances and thicknesses are much too random to be modeled as periodic, and there are additional random components on top of this. If the spacing between the warps and wefts were more deterministically repeatable (like the deterministic spacing of the golf ball dimples in Figure 1(d)), then this periodic component might be modeled as a legitimate profile mean function and handled via existing profile monitoring methods. Nonetheless, the random nature of the spacing precludes this approach.

Theoretically, the joint distribution of the collection of pixels in a stochastic textured surface sample provides a complete statistical representation, including capturing any stochastic spatial dynamics. Direct estimation of such a high-dimensional nonparametric distribution is obviously infeasible. However, under the stationary Markov random field assumptions of Section 2, the joint distribution can be implicitly and approximately characterized by the conditional distribution of individual element/pixel values in the image, given the values of a collection of neighboring pixels, with the conditional distribution estimated using supervised learning methods applied to an in-control training image sample(s). This technique was used in Bostanabad et al. (2016) to characterize and reconstruct binary material microstructure images. In this work we use this supervised learning approach to obtain an implicit representation of the in-control (i.e., normal) statistical behavior of a stochastic textured surface. Our objective is to develop a statistical monitoring approach for detecting local phenomena or defects in the manufactured stochastic textured surfaces that are statistically inconsistent with the in-control behavior, as represented by the implicit supervised learning characterization.

There is a growing body of work on image monitoring (see, e.g., the review paper of Megahed, Woodall, and Camelio, 2011). Some early works directly monitored the intensity levels of the pixels in the images (Jiang and Jiang 1998, Armingol et al. 2003). Most of later methods first extracted a small set of predefined feature characteristics from the images and then monitored directly those specific characteristics or the statistics obtained from them. Common characteristics include length, width, and area (Tan, Chang, and Hsieh 1996), shape (Liang and Chiou 2008), and diameter (Lyu and Chen 2009) of specific features identified in the image. Other work has monitored frequency domain characteristics based on wavelets (Liu and MacGregor 2006, Lin 2007a, Lin 2007b) and frequency spectrum features (Tunák, Linka and Volf 2009), principle components (Bharati and MacGregor 1998, Bharati, MacGregor and Tropper 2003), and grey level co-occurrence matrix features (Tunák and Linka 2008). Recently, Megahed et al. (2012) monitor summary statistics comprised of the average intensity levels of predefined windows of various sizes across the images.

Using predefined features is problem-specific, by definition, and requires that the users have a fairly specific idea of the nature of the defects that they would like to detect. Our goal is to develop a more general approach that can detect general local deviations from the normal in-control statistical behavior of stochastic textured surfaces, where the normal in-control statistical behavior is modeled in a reasonably generic manner from the in-control training images. In this regard, the approach is analogous to the classic Shewhart control charting approach (Montgomery 2009), which generically characterizes in-control behavior from a training sample (Phase I), and monitors future samples to detect general departures from the in-control behavior (Phase II). Our primary monitoring statistic is derived from some appropriate spatial moving statistic (to be defined in Section 3), computed from the residual errors of the supervised learning model that characterizes the stochastic textured surface of interest. In addition to monitoring, our approach is designed to help users diagnose the cause of the deviations from normal behavior via highlighting pixels with large spatial moving statistics.

To the best of our knowledge, most industrial machine vision algorithms are intended for situations in which there is a legitimate gold-standard image and/or there are distinct predefined features (e.g., edges, corners, circles, spectral peak frequencies/amplitudes, average intensity levels, etc.) that can be detected with standard image processing toolboxes. The types of stochastic textured surfaces to which this work applies have neither a gold standard image nor standard features that can be detected. The main contribution of this work is developing an approach that can be used for this general class of stochastic textured surface inspection images, for which there is currently a hole in the existing literature.

It should be noted that, although our algorithm could be applied to monitoring surfaces with deterministically repeating patterns like the dimpled surface in Figure 1(d), we do not recommend it for that. A much more sensible approach would take into account the known, deterministic spacing and size of the dimples to either (i) compare inspection images to a gold standard dimpled surface representing the nominal geometry (after proper registration and alignment of the images) or (ii) extract relevant features related to the dimple size and spacing and monitor the features.

The remainder of the paper is organized as follows. Section 2 describes how supervised learning can be used to implicitly characterize the normal in-control spatial statistical behavior of the stochastic textured surfaces. Section 3 introduces two spatial moving statistics and the primary monitoring statistic. Section 4 elaborates details of our monitoring and diagnostic algorithm. Section 5 and Section 6 illustrate and compare the approach with three other approaches in a simulation study and in the textile example depicted in Figure 1(a), respectively. Section 7 concludes this paper.

## 2. Modeling the Spatial Statistical Characteristics of the Stochastic Textured Surfaces via Supervised Learning

Suppose an image is comprised of $M$ pixels, and let $\mathbf{Y}_j = [y_{j,1}, \ y_{j,2}, \ \ldots, y_{j,M}\ ]^T$ ($j$ = 1, 2, . . ., $N$) denote the set of ordered pixels for the $j^{\text{th}}$ image in a sample of $N$ images. We use the subscript $j$ later for indexing images; however, we will often omit it for simplicity, unless necessary. Suppose the elements of $\mathbf{Y}$ are ordered in a row raster scan pixel sequence of left-to-right, moving from the top row to the bottom row of the image, as illustrated in Figure 2. Let $f(\mathbf{Y})$ denote the joint distribution of $\mathbf{Y}$, which theoretically provides the most complete characterization of the statistical behavior of the stochastic textured surface. However, it is clearly infeasible to estimate such high-dimensional nonparametric distributions directly. In light of this, consider the factorization:

$$f(\mathbf{Y}) = f(y_M | y_{M-1}, y_{M-2}, \ldots) f(y_{M-1} | y_{M-2}, y_{M-3}, \ldots) \ldots f(y_2 | y_1) f(y_1) = \prod_{i=1}^{M} f\left(y_i \middle| \mathbf{Y}^{(i)}\right),$$

where $\mathbf{Y}^{(i)} = \{y_k\text{: } k = 1, \ldots, i-1\}$. The notation is illustrated in Figure 2.

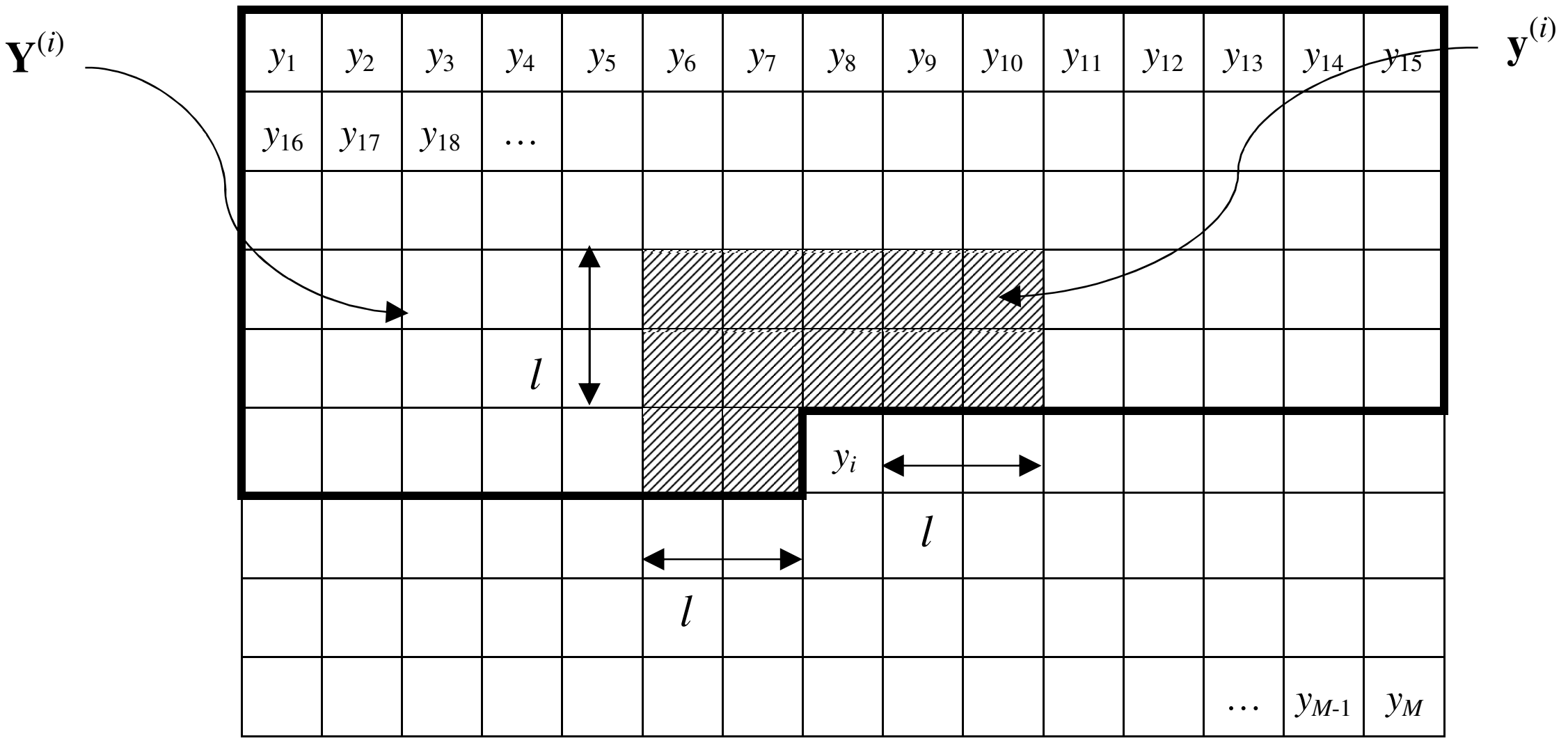

**Figure 2**. Illustration of the notation with a stylized pixelated image (each cell represents a pixel). The pixels inside the area with bold borderlines are the elements of $\mathbf{Y}^{(i)}$, and the shaded pixels are the elements of $\mathbf{y}^{(i)}$. Given $\mathbf{y}^{(i)}$, $y_i$ is assumed independent of $\mathbf{Y}^{(i)} \backslash \mathbf{y}^{(i)}$.

Using this factorization, we can implicitly obtain the joint distribution $f(\mathbf{Y})$ by learning each conditional distribution $f(y_i|\mathbf{Y}^{(i)})$ via fitting some appropriate supervised learning model to predict the "response" variable $y_i$ as a function of the set of "predictor" variables $\mathbf{Y}^{(i)}$. Without further assumptions, this is unmanageable, in part because it would require learning $M$ separate models, and many of them have an extremely high-dimensional predictor space (e.g., $\mathbf{Y}^{(M)}$ is $(M-1)$-dimensional). To make the problem more manageable, we assume the following Markov random field (MRF) properties for the stochastic textured surfaces, which are generally quite reasonable and are often assumed in texture synthesis problems (Efros and Leung 1999, Levina and Bickel 2006). The first MRF property is *locality:* there exists a neighborhood $\mathbf{y}^{(i)} = \{y_k \in \mathbf{Y}^{(i)}$: pixel $k$ is within some neighborhood of pixel $i\}$ such that given $\mathbf{y}^{(i)}$, $y_i$ is independent of all other pixels in $\mathbf{Y}^{(i)}\backslash\mathbf{y}^{(i)}$, i.e., such that $f(y_i \mid \mathbf{Y}^{(i)}) = f(y_i \mid \mathbf{y}^{(i)})$. Figure 2 depicts this neighborhood $\mathbf{y}^{(i)}$ as the shaded region. The second MRF property is s*tationarity*: $f(y_i = y \mid \mathbf{y}^{(i)} = \mathbf{y})$, as a function of $y$ and $\mathbf{y}$, is independent of pixel location $i$.

By the locality assumption,

$$f(\mathbf{Y}) = \prod_{i=1}^{M} f\left(y_i \middle| \mathbf{Y}^{(i)}\right) = \prod_{i=1}^{M} f\left(y_i \middle| \mathbf{y}^{(i)}\right). \tag{1}$$

Thus, we can obtain $f(\mathbf{Y})$ by learning $f(y_i|\mathbf{y}^{(i)})$, which is more computationally feasible since the size of $\mathbf{y}^{(i)}$ is much smaller than that of $\mathbf{Y}^{(i)}$. The stationarity assumption enables us to estimate $f(y_i|\mathbf{y}^{(i)})$ by fitting an appropriate supervised learning model to a set of training data constructed from the collection of pixels in some training image $\mathbf{Y}$. The training data array consists of $M$ rows with each row corresponding to one of the pixels in $\mathbf{Y}$. The $i^{\text{th}}$ row of the training data set is comprised of $\{y_i, \mathbf{y}^{(i)}\}$, where $y_i$ and $\mathbf{y}^{(i)}$ represent the response and predictor variables, respectively. When fitting the supervised learning model, the first column is treated as the response column, and the remaining columns as the predictor columns. Henceforth, $M$ denotes the number of pixels in the interior of the image, excluding a small boundary region just large enough that the first pixel $y_1$ has a full size neighborhood $\mathbf{y}^{(1)}$.

If the greyscale pixel values were coarsely discretized, then the conditional distribution of $y_i \mid$

$\mathbf{y}^{(i)}$ would be multinomial, and any appropriate supervised learning classifier could be used to learn the multinomial probabilities as a function of the predictor variables $\mathbf{y}^{(i)}$. For the case of binary images representing two-phase material microstructure samples, Bostanabad et al. (2016) used this supervised learning approach to learn the Bernoulli conditional probabilities of $y_i \mid \mathbf{y}^{(i)}$. Their fitted supervised learning model provided an implicit characterization (via (1)) of the microstructure, which they used to reconstruct microstructure samples that were statistically equivalent to the original training sample.

Because we are assuming finely discretized greyscale intensity levels, we treat them as continuous and consider a supervised learning model of the general form $y_i = g(\mathbf{y}^{(i)}) + \varepsilon_i$, where $g(\mathbf{y}^{(i)})$ is the mean of the conditional distribution $f(y_i|\mathbf{y}^{(i)})$, and $\varepsilon_i$ is a zero-mean error. Applying an off-the-shelf supervised learning algorithm to an in-control image, we obtain a model that represents the estimated conditional mean function $\hat{g}(\mathbf{y}^{(i)})$. Although the conditional mean does not fully represent the conditional distribution, it does provide rich enough information to monitor for deviations from the in-control statistical behavior of the stochastic textured surfaces. As will be discussed in Section 3, we use the residuals of the supervised learning model for our monitoring and diagnostic purposes.

It should be noted that other ways of ordering the pixels, such as the zigzag scanning method used in Megahed and Camelio (2012), could result in a different fitted supervised learning model, especially if the surface is not isotropic. If desired, one could use cross-validation to select the best ordering as the one that minimizes the cross-validation error sum of squares. In all of our examples, we only considered the raster scan order depicted in Figure 2.

## 3. Choice of Monitoring Statistic

In this section, we develop our monitoring statistics that are based on the residuals of the in-control supervised learning model. Section 3.1 presents the monitoring approach in terms of a general spatial moving statistic (SMS) that appropriately aggregates the local residual behavior, and Sections 3.2 and 3.3 discusses two specific statistics to serve as the SMS.

### 3.1 Monitoring based on local residual behaviors

Henceforth, let $g(\mathbf{y}^{(i)})$ denote the conditional mean model fitted to a training image(s) that are known to represent in-control behavior. For a new inspection image, denote the residual for the $i^{\text{th}}$ pixel ($i$ = 1, 2, . . ., $M$) by

$$r_i = y_i - \hat{g}(\mathbf{y}^{(i)}). \quad (2)$$

Note that the residuals themselves constitute an image that corresponds pixel-wise to the image from which the residuals are computed (e.g., see Figure 3). If the new image also behaves as under the in-control conditions, then the residuals $\{r_i : i = 1, 2, \ldots, M\}$ should behave approximately as white noise, although departures from white noise are automatically adjusted for, via the way the control limits are determined in our approach (see Section 4.1). In contrast, if the new image has defects or other departures from the in-control stochastic behavior, then the residuals should behave differently than that when the image is in-control. Hence, our monitoring procedure is based on monitoring the residuals in a manner to be described shortly (see the online supplement for this paper for further discussion on the types of defects that our algorithm can detect, which are reasonably general).

Monitoring individual residuals may not be sensitive enough to detect milder defects, for the same reason that Shewhart individual charts are not sensitive enough to small mean shifts. Consequently, we use moving window to aggregate the residuals in some manner over an appropriately sized spatial neighborhood of the image. To measure the degree of deviation from in-control behavior of the residuals within the moving window neighborhood, we use a SMS that is some statistic computed from the residuals within a spatial moving window that is scanned across the residual image. Let $w$ denote the width (in number of pixels) of the square spatial moving window, which contains $n = w^2$ residuals. For example, the window with the bold border in Figure 3 depicts the moving window centered at the $i^{\text{th}}$ pixel. The SMS at the $i^{\text{th}}$ pixel of the $j^{\text{th}}$ image, denoted by $SMS_{j,i}$, is defined as some function of the $w^2$ residuals within the moving window surrounding the $i^{\text{th}}$ pixel of the $j^{\text{th}}$ image. In this paper, we consider two such SMSs that are intuitively appealing and that we have found to result in good defect detection performance in our examples: (i) a one-sample Anderson–Darling (A-D) statistic and (ii) a Box–Pierce (B-P) type statistic, which we describe in Sections 3.2 and Section 3.3, respectively.

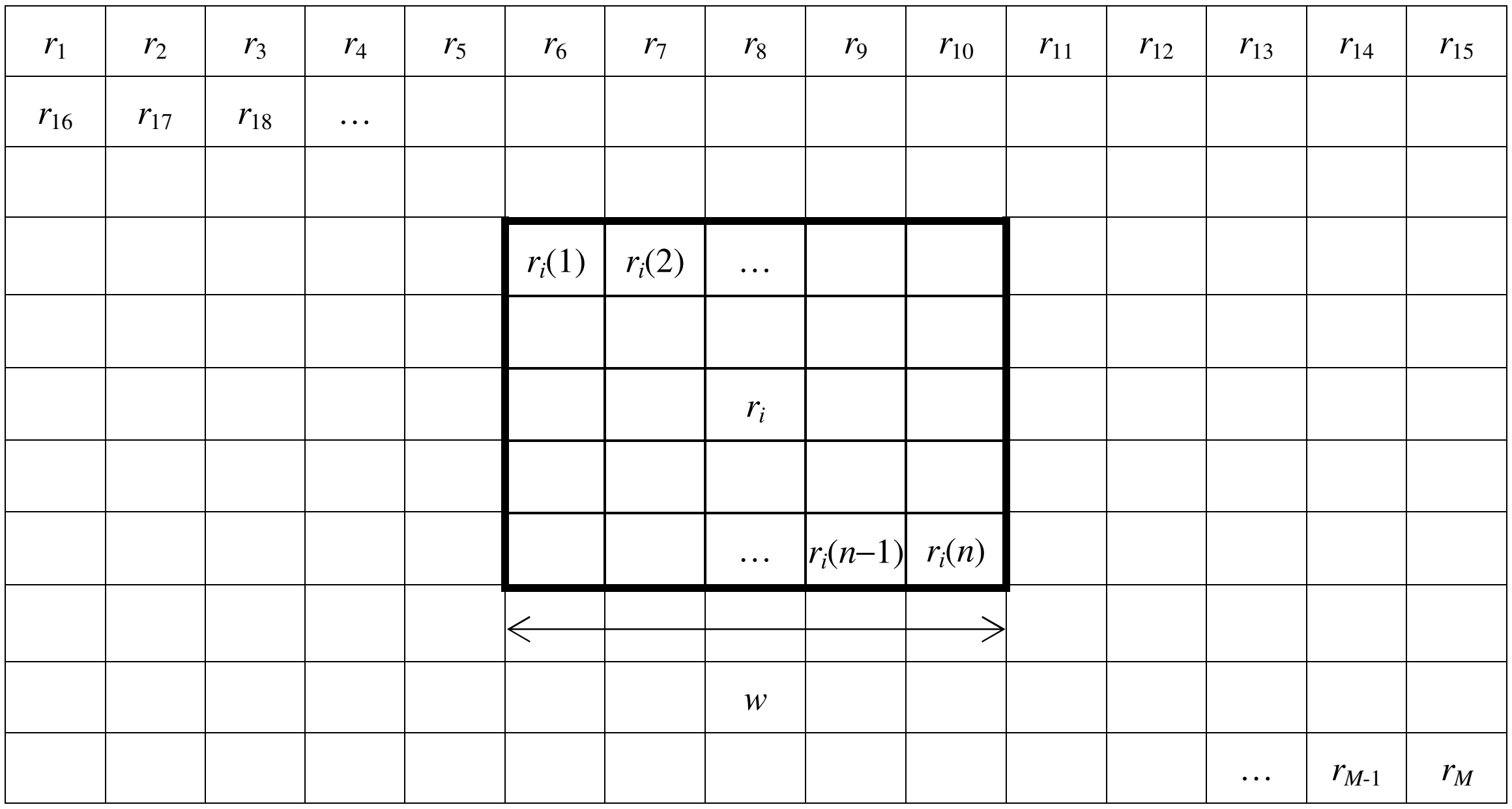

**Figure 3**. An image of residuals illustrating the spatial moving windows: each cell corresponds to a pixel of the image from which the residuals are computed. The pixels $\{r_i(1), r_i(2), \ldots, r_i(n)\}$ inside the bold lines are the elements of the square moving window of $n = w^2$ residuals centered at the $i^{\text{th}}$ pixel, corresponding to residual $r_i \equiv r_i((w^2 + 1)/2)$.

Our algorithm is intended for monitoring and diagnosing individual images using a single aggregate summary statistic for each image. Moreover, the intent is that an alarm will be sounded if an individual image contains a defect, as opposed to requiring that defects persistently occur across a consecutive set of images. In this respect, our approach is akin to a Shewhart individual chart. We define our monitoring statistic for the $j^{\text{th}}$ image to be

$$S_j = \max_{i=1 \ldots M} SMS_{j,i} \quad (3)$$

If the defects do occur persistently across consecutive images, then our monitoring approach could be enhanced by using a EWMA-type or CUSUM-type accumulation of $S_j$, although we do not pursue this in this paper.

### 3.2 A-D statistic

As discussed in Section 3.1, we expect a local change in the distribution of the residuals in the defect region, relative to the in-control residual distribution. We represent the latter by a reference cumulative density function (cdf), denoted by $\varphi$, of all the residuals **R** computed from a representative in-control image(s). As a statistic that measures the deviation (from $\varphi$) of the residual distribution within some neighborhood of a pixel, we consider a one-sample A-D statistic (Anderson and Darling 1954), which compares the empirical cdf of the residuals within a spatial moving window versus $\varphi$. We also considered a one sample Kolmogorov-Smirnov statistic, but do not pursue it here, because we found that the A-D statistic performed better. This perhaps was because the A-D statistic is more sensitive to changes in the tails of the distribution, which correspond to large-magnitude residuals.

Let the residuals $\{r_i(1), r_i(2), \ldots, r_i(n)\}$ within the moving window around the $i^{\text{th}}$ pixel be ordered from smallest to largest. The one-sample A-D SMS at the $i^{\text{th}}$ pixel is defined as:

$$A_i^2 = -n - \sum_{k=1}^{n} \frac{2k-1}{n} \ln\{\varphi(r_i(k))[1 - \varphi(r_i(n+1-k))]\} \tag{4}$$

Since the sample size for the training image is quite large, one might consider using the empirical cdf of the residuals **R** (denoted by $F$, for which a corresponding histogram is shown in Figure 4) for the training image as $\varphi$ in (4). However, this causes a potential problem, because the one-sample A-D statistic is infinite/undefined if any of the $n$ elements within the moving window are beyond the support of $\varphi$. To illustrate, Figure 4 shows a histogram of approximately 0.25 million residuals from a training image in one of our examples, the support of which extends from [−2.45, 2.84]. Thus, if a new image contains a residual that falls outside the interval [−2.45, 2.84], which happened frequently in our example (even with the process was in-control), the statistic in (4) is infinite for any moving window containing that residual.

To avoid this problem, instead of using $\varphi = F$ directly, we replace its upper and lower tails with an exponentially decaying tail approximation. The upper tail approximation for $r > 2.38$ is illustrated in Figure 4. More specifically, let $r_{q_l}$ and $r_{1-q_u}$ denote the lower $q_l$ quantile and upper $q_u$ quantile of $F$ for some small probabilities $q_l$ and $q_u$ such that $F(r_{q_l}) = q_l$ and $F(r_{1-q_u}) = 1 - q_u$. The probabilities $q_l$ and $q_u$ should be large enough to have enough tail observations to get a good estimate of the exponential rate parameters for the tail approximation, but otherwise as small as possible. For our examples we have used values $q_l \approx q_u \approx 1.6 \times 10^{-3}$, which, because of the large number of pixels in typical images, translate to around 400 observations in each tail. To estimate the rate parameters, we fit the observations corresponding to the lower and upper tails of $F$ with the exponential probability density functions (pdfs):

$$f(r) = \begin{cases} \frac{q_l}{\lambda_l} exp\left\{\frac{r - r_{q_l}}{\lambda_l}\right\} & : \quad r \le r_{q_l} \\ \frac{q_u}{\lambda_u} exp\left\{-\frac{r - r_{1-q_u}}{\lambda_u}\right\} & : \quad r \ge r_{1-q_u} \end{cases}$$

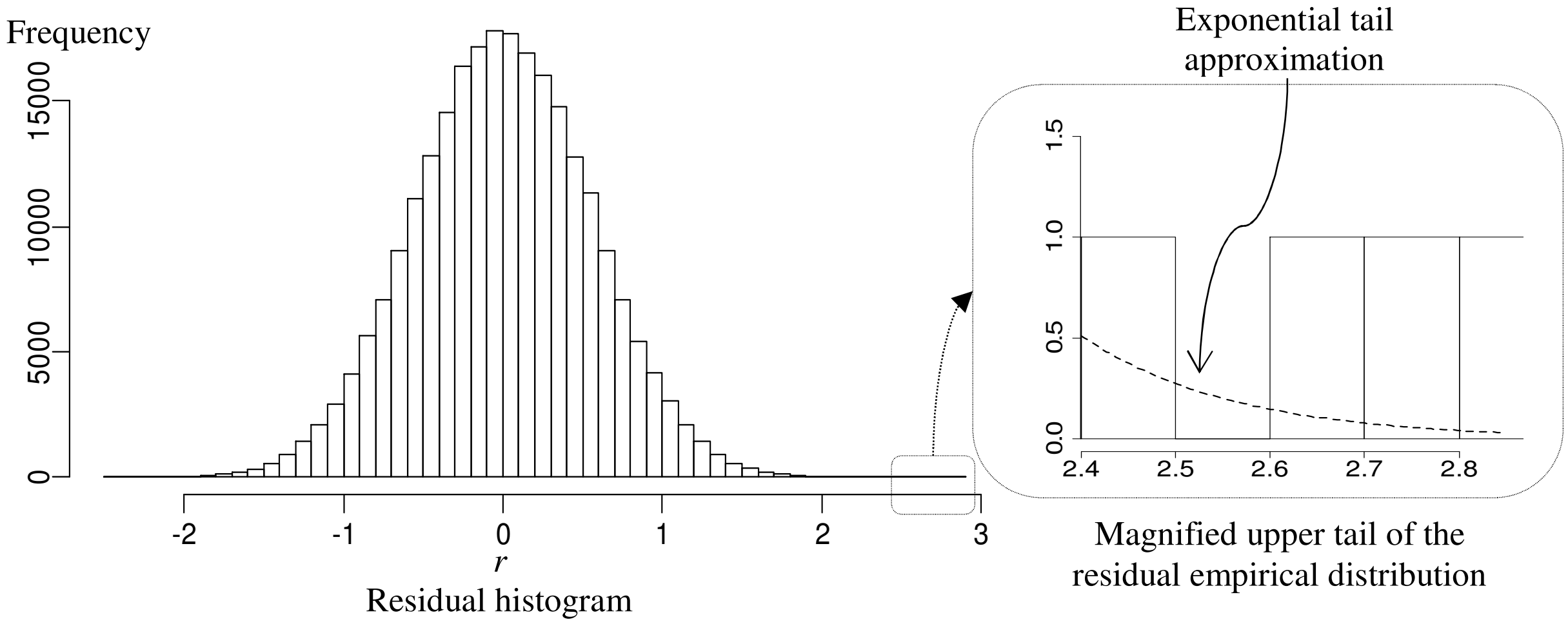

**Figure 4**. Approximating the upper and lower tails of the residual empirical cdf with an exponentially decaying distribution. The sample size is approximately 0.25 million pixels.

The maximum likelihood estimators of the lower and upper exponential rate parameters are $\lambda_l = r_{q_l} - ave\{r_i: r_i \le r_{q_l}\}$, and $\lambda_u = ave\{r_i: r_i \ge r_{1-q_u}\} - r_{1-q_u}$. We then choose a very small probability $p$ ($p = 5/M$ in the example in Figure 4) and replace $F(r)$ by its exponential tail approximation for $r < r_p$ and $r > r_{1-p}$, where $r_p$ and $r_{1-p}$ are the lower and upper $p$ quantiles of $F$. That is, for $\varphi$ in (4) we use

$$\varphi(r) = \begin{cases} p \times exp\left\{\frac{r-r_p}{\lambda_l}\right\}: & r \le r_p \\ F(r): & r_p < r < r_{1-p} \\ 1 - p \times exp\left\{-\frac{r-r_{1-p}}{\lambda_u}\right\}: & r \ge r_{1-p} \end{cases}$$

### 3.3 B-P type statistic

A B-P (aka portmanteau) test (Box and Pierce 1970) is widely used for testing the existence of autocorrelations in time series. Likewise, a B-P type statistic can be used to detect spatial correlations in our stochastic textured surface images. Because local defects in the stochastic textured surfaces are likely to result in local spatial correlations in the residuals, the B-P type statistic is intuitively appealing for our objective. We define the B-P type SMS for the $i^{\text{th}}$ pixel as

$$T_i = \sum_{k=1}^{n} \widehat{Cov}_{i,k}^2 \tag{5}$$

where $\widehat{Cov}_{i,k}^2$ is some local estimate of the covariance between the residual $r_i$ at the $i^{\text{th}}$ pixel and another residual $r_k$ within the moving window of *n* pixels surrounding the $i^{\text{th}}$ pixel (e.g., the moving window in Figure 3). Note that $\widehat{Cov}_{i,i}^2$ is included in $T_i$ in (5). To estimate $\widehat{Cov}_{i,k}$, we use a kernel weighted window centered at the $i^{\text{th}}$ pixel. For ease of illustration, let $i_1$ and $i_2$ be the row and column indices of the $i^{\text{th}}$ pixel, and let $k_1$ and $k_2$ be the row and column indices of the $k^{\text{th}}$ pixel. Then,

$$\widehat{Cov}_{i,k} = \frac{\sum_{h=-\infty}^{\infty}\sum_{m=-\infty}^{\infty} K(h,m) r_{i_1-h,i_2-m} r_{k_1-h,k_2-m}}{\sum_{h=-\infty}^{\infty}\sum_{m=-\infty}^{\infty} K(h,m)}$$

where $K(h,m)$ is the Epanechnikov quadratic kernel:

$$K(h,m) = \begin{cases} \frac{3}{4}\left(1 - \frac{h^2+m^2}{\left(\frac{w+1}{2}\right)^2}\right): & h^2 + m^2 \le \left(\frac{w+1}{2}\right)^2 \\ 0 & : \quad otherwise \end{cases} \tag{6}$$

## 4. Stochastic Textured Surface Monitoring and Diagnostic Algorithm

Our approach involves two stages: monitoring and diagnosis. The monitoring stage has two phases. The first is an offline training phase (Phase I) that constructs a control limit based on the empirical distribution of the monitoring statistic computed for a set of in-control images. As defined in (3), the monitoring statistic for the $j^{\text{th}}$ image is the maximum of the SMS values across all pixels in that image, where our SMS for pixel $i$ in image $j$ is either $A_i^2$ or $T_i$ for image $j$. The SMS values are calculated from the residuals of the supervised learning model that characterizes the stochastic textured surface to be monitored. The second phase is an online monitoring phase (Phase II) that computes a monitoring statistic for each new image similarly to that in Phase I. If the monitoring statistic is beyond a control limit, an alarm is sounded, and the diagnostic stage is invoked. The diagnostic stage constructs a binary image that corresponds pixel-to-pixel with the original image and highlights the pixels with SMS values larger than some threshold. We refer to these as *diagnostic images*. We provide details of Phase I of the monitoring stage in Section 4.1. Phase II of the monitoring stage, as well as the diagnostic stage, are discussed in Section 4.2.

### 4.1 Phase I of the Monitoring Stage: Fitting the Supervised Learning Model and Establishing the Control Limits

Phase I of the monitoring stage begins with choosing a region from an image (or images) that is known to be in-control, from which to construct the training data set for fitting the supervised learning model as described in Section 2. The neighborhood structure must also be chosen. This can be flexible, but to simplify the discussion, we define the neighborhood by a single parameter $l$, which is the number of pixels to the right/left and above the response pixel, corresponding to our raster scan method. Figure 2 illustrates such a neighborhood with $l = 2$. The value of $l$ should be large enough to include all important predictor variables (such that the MRF locality property holds), but not so large as to incur unnecessary computational expense. We recommend choosing $l$ via cross-validation during the process of fitting the supervised learning model to minimize some measure of cross-validation error. Readers are referred to Bostanabad et al. (2016) for further details of the model fitting procedure.

After fitting the supervised learning model, we apply it to a new image $j$ (that is believed to be in-control) to calculate the predictions $\left\{\hat{g}\left(\mathbf{y}_j^{(i)}\right): i = 1, 2, \ldots\right\}$ and the corresponding residual errors $\{r_{j,i}: i = 1, 2, \ldots\}$ via (2). After that, the SMS values $\{SMS_{j,i}: i = 1, 2, \ldots\}$ are calculated as described in Section 3. Finally, the monitoring statistic $S_j$ for the $j^{th}$ image is computed via (3). This process is repeated for a set of $N$ in-control images (i.e., for $j = 1, 2, \ldots, N$) to give a sample $\{S_j: j = 1, 2, \ldots, N\}$ of monitoring statistics that represent the in-control state. Given the complexity of the supervised learning model and the image texture characteristics, it is not possible to derive some exact (or even reasonably approximate) theoretical distribution of the residuals and the resulting theoretical distribution of the monitoring statistic $S$ in order to set the control limits. Thus, we set the control limits based on the empirical distribution of $\{S_j: j = 1, 2, \ldots, N\}$ from the set of in-control Phase I images, which is often available in practice.

Figure 5 shows histograms of $\{S_j: j = 1, 2, \ldots, N\}$ for the $N = 1{,}000$ Phase I images for the example in Section 5 based on A-D (Figure 5(a)) and B-P type (Figure 5(b)) SMSs, respectively. The theoretical support of the distribution of $S$ in either case is $[0, \infty)$, and a larger $S_j$ indicates a higher likelihood that image $j$ contains a defect. Thus, there is only an upper control limit. Letting $\alpha$ (e.g. $\alpha = 0.003$) denote the desired Type I error for an individual $j^{th}$ image, we set the control limit as the $(1 - \alpha)$ quantile of the empirical distribution of $\{S_j: j = 1, 2, \ldots, N\}$.

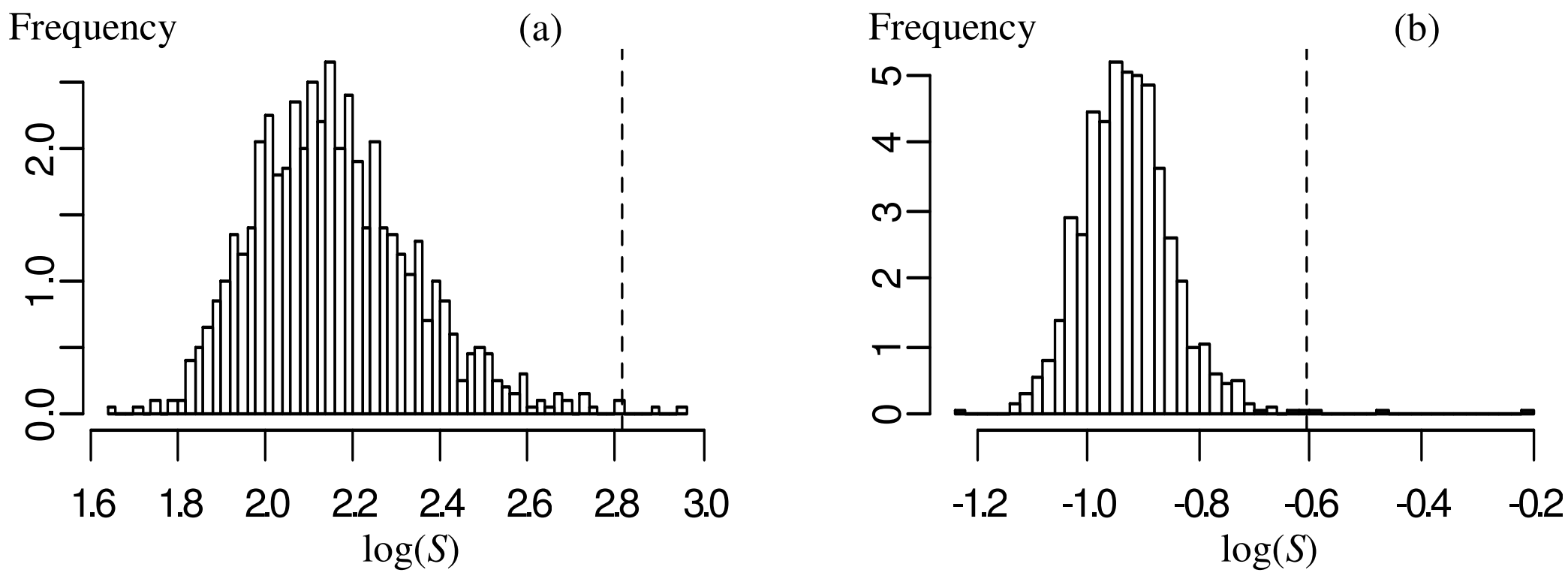

**Figure 5**. Histograms of Phase I $\{S_j: j = 1, 2, \ldots, N\}$ in log scale based on (a) the A-D statistic and (b) the B-P type statistic, computed from $N = 1{,}000$ in-control images for the example in Section 5 with $w = 25$. The dashed lines are the control limits corresponding to $\alpha = 0.003$.

### 4.2 Phase II of the Monitoring Stage and the Diagnostic Stage

Phase II of the monitoring stage, which involves many of the same calculations as Phase I, and the diagnostic stage are relatively straightforward. First, a data array is constructed from each new image as described in Section 2, and the supervised learning model fitted in Phase I is applied to generate the predictions $\hat{g}(\mathbf{y}^{(i)})$ and the corresponding residuals for the new image. The SMS values at each pixel of the new image are computed from these residuals, and the monitoring statistic $S$ is computed via (3), after which it is compared to the control limit calculated in Phase I.

If an alarm is sounded for an image, the algorithm invokes the diagnostic stage, which compares the SMS values at all pixels in that image with a diagnostic threshold. A binary diagnostic image is then constructed by plotting every pixel with SMS value larger than the diagnostic threshold as a black pixel, and the remaining pixels in the image as white pixels. The diagnostic threshold has no connection to the control limit, as the former applies to the SMS values, whereas the latter applies to the $S$ values. Moreover, while the control limit is chosen to control the Type I error, the diagnostic threshold is chosen purely to facilitate visualization of the nature of the defect. Our recommended strategy for selecting the diagnostic threshold is so that it results in a small but acceptable level of noise (i.e., black pixels that are not associated with actual defects) in the diagnostic image. We accomplish this by setting the diagnostic threshold at the $(1 - n_D/M_{SMS})$ quantile of the empirical distribution of all SMS values computed for all Phase I images, where $M_{SMS}$ is the number of SMS values computed for each Phase I image, and $n_D$ is the desired average number of black (noise) pixels in a diagnostic image of an in-control image. The choice of $n_D$ depends on the image size in general. We have used $n_D$ ~ 5—10 for an image size of 250×250 pixels in our examples. Alternatively, instead of selecting a single $n_D$, users could vary $n_D$ as the diagnostic image is dynamically changed to better facilitate visualization of the defect.

## 5. Simulation Study

In this section, we demonstrate and evaluate our approach with simulated images of the 2-D stochastic process depicted in Figure 1(b). We also compare its performance with three alternative methods. The images were generated via the spatial autoregressive model $y(i, k) =$ $\phi_1 y(i-1,k) + \phi_2 y(i,k-1) + \varepsilon(i,k)$, where $y(i, k)$ denotes the image greyscale level at pixel location $(i, k)$ with $i$ and $k$ the row and column indices, respectively. We used $\phi_1 = 0.6$, $\phi_2 = 0.35$, and $\varepsilon$ a zero-mean Gaussian white noise. After generating the process, we translated/rescaled it to the interval [0, 255] to obtain the corresponding greyscale image for plotting purposes. For applying our algorithm, all images were subsequently standardized by subtracting from each pixel the average greyscale value for all pixels in that image and then dividing by the greyscale standard deviation for all pixels in the image. For real examples, this is helpful if the lighting conditions vary from image to image, although ideally the lighting should be controlled. Note that the MRF assumptions hold for the images in this example, by construction.

### 5.1 Monitoring stage

We evaluated the monitoring performance of our algorithm with 10 replicates of the following experiment. On each replicate we first generated an image of size 500×500 pixels, similar to the one in Figure 1(b), for model fitting (discussed in Section 4.1). Then, we used a regression tree as the supervised learning model (any appropriate supervised learner could be used) because it is more computationally reasonable to fit for large training data sets. The neighborhood size $l$ was obtained during the tree fitting process as the one that minimized the cross-validation sum-of-squares error. This resulted in $l$ = 1, which agrees with the lag-one autoregressive model used to generate the data. For real examples, like the textile application in Section 6, the cross-validation procedure will typically select a much larger value of $l$.

To construct the control limit, we generated a Phase I set of $N$ = 1000 in-control images, each of size 250×250 pixels, in the same manner as the training image used for model fitting. Using the fitted regression tree from the training image, for each image $j$ in the Phase I set, we computed the SMS values for every pixel and then the monitoring statistic $S_j$ in (3), as described in Section 4.1. We considered both the A-D and B-P type SMSs, each with several spatial moving window sizes ($w$ = 5, 15, and 25), for comparison purposes. For the A-D statistic, we chose $q_l \approx q_u \approx 1.6 \times 10^{-3}$ in order to give around 400 observations in each tail for estimating the exponential tail parameters. We also chose $p \approx 2 \times 10^{-5}$, for which $r_p = -2.16$ and $r_{1-p} = 2.38$, and we replaced $F(r)$ by its exponential tail approximation for $r \notin (r_p, r_{1-p})$. From the empirical distribution of $\{S_j: j = 1, 2, . . ., 1000\}$ (the histograms for which are shown in Figure 5 for $w$ = 25) and with a desired Type I error rate of $\alpha = 0.003$, we selected the control limit, which depended on the choice of $w$.

For monitoring performance evaluation, we generated 400 Phase II images, all containing defects, by first generating an in-control image of size 250×250 pixels from the same spatial autoregressive model and then creating a defect and superimposing on the image. We considered "white noise defects" that were Gaussian white noise process (i.e., the spatial autoregressive process with $\phi_1 = \phi_2 = 0$) with the same mean and standard deviation as the white noise $\varepsilon$ in the in-control process. The defect regions that we superimposed were ellipsoidal shaped and of sizes 5×5, 5×21, 9×21, and 15×21 (the sizes refer to the lengths of the major and minor axes of the ellipses, which were aligned with the horizontal and vertical axes of the images). Randomly positioned defects of each these sizes were added to 100 images each (one defect added to each image) to generate the Phase II out-of-control images. The first row of Figure 6 shows some examples of these Phase II images, the defects of which are difficult to spot visually.

Table 1 reports the average power across 10 replicates for our approach (with different combinations of the SMS statistic and $w$) for all four defect sizes mentioned above. Notice that the algorithm generally detects defects with higher power when $w$ is approximately the same size as the defects (we have also observed this phenomenon in other examples). This is intuitively reasonable, because the SMS (either A-D or B-P type) is larger when its window contains more pixels in the defect region and fewer pixels in the normal region. However, this is not an implementable guideline for choosing $w$, because defect sizes may not be known in advance. Regarding choice of $w$, we have observed that for the A-D-based statistic, the performance suffers more when $w$ is larger than the defects than it does when $w$ is smaller than the defects. This can be observed by comparing the three columns for the A-D-based statistic in Table 1.

Consequently, for the A-D-based statistic, we recommend choosing $w$ to be approximately the same as the smallest defect size of interest.

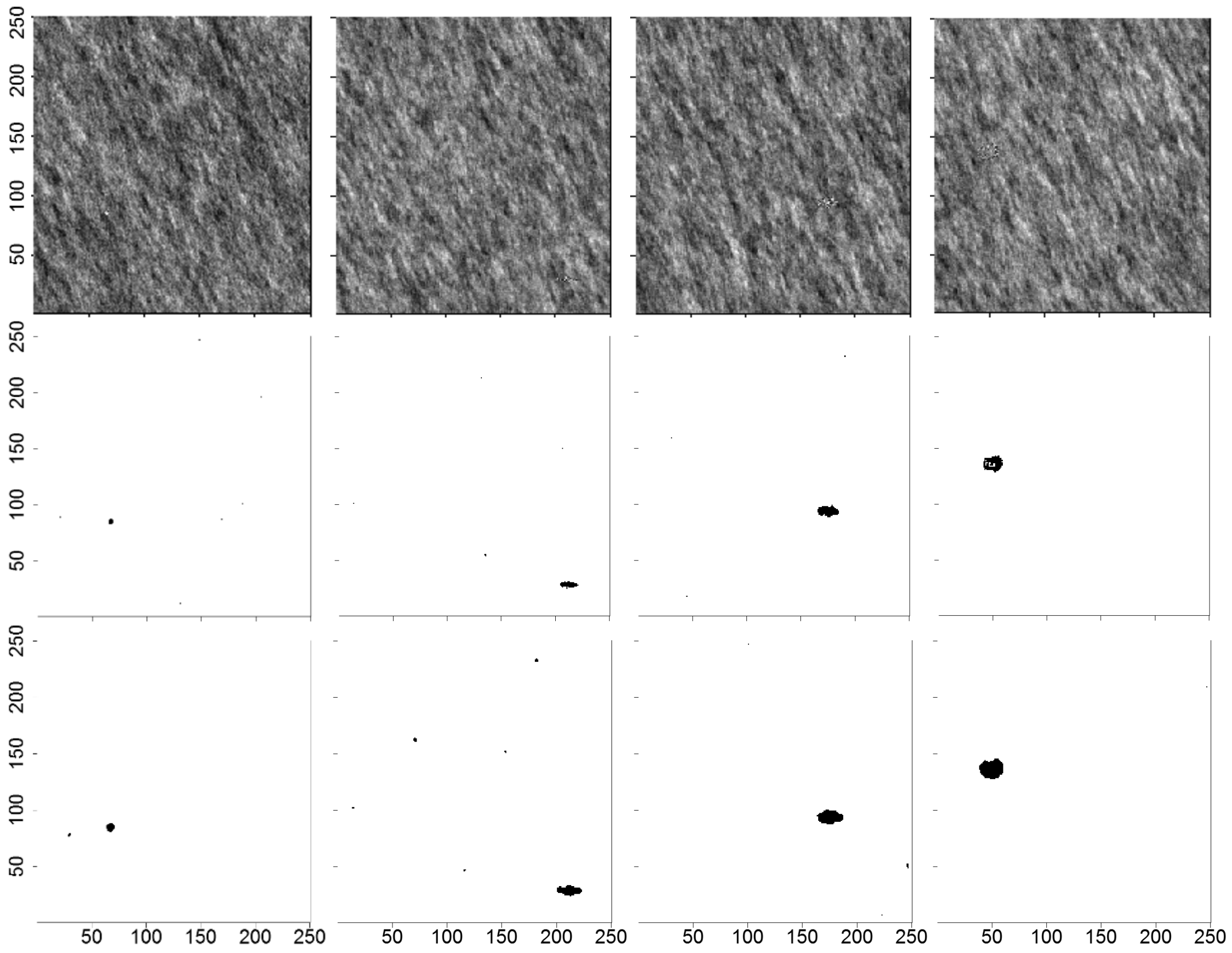

**Figure 6**. Phase II images in the simulation example containing white noise defects (top row) and their diagnostic images using the A-D-based statistic (middle row) and the B-P-type statistic (bottom row). The defects in the images in the top panels have different sizes: (1st column) 5×5, (2nd column) 5×21, (3rd column) 9×21, and (4th column) 15×21.

**Table 1**. Average powers in 10 replicates of our approach at $\alpha = 0.003$

| Defect sizes | A-D | | | B-P | | |
|---|---|---|---|---|---|---|
| | $w = 5$ | $w = 15$ | $w = 25$ | $w = 5$ | $w = 15$ | $w = 25$ |
| 5×5 | 0.205 | 0.004 | 0.003 | 0.955 | 0.884 | 0.858 |
| 5×21 | 0.785 | 0.791 | 0.247 | 0.997 | 1.000 | 1.000 |
| 9×21 | 0.964 | 1.000 | 0.987 | 1.000 | 1.000 | 1.000 |
| 15×21 | 0.990 | 1.000 | 1.000 | 1.000 | 1.000 | 1.000 |

For the B-P-type statistic, the monitoring performance for all but the smallest defects was almost perfect even when $w$ is larger than the defect size, as can be seen from the three columns for the B-P-type statistic in Table 1. To demonstrate the extent to which the monitoring statistics in these cases exceed the control limits, Figure 7(a) shows boxplots of $(\bar{S}_w - CL_w)/(UCL_w - CL_w)$ across the 10 replicates, where $\bar{S}_w$ is the average B-P-type monitoring statistic using an SMS size of $w$ for all Phase II images containing defects of size 5×21. Figures 7(b) and 7(c) show similar boxplots, but for defect sizes 9×21 and 15×21, respectively. By comparing the boxplots in each panel of Figure 7 we see that as $w$ increases, the monitoring statistic tends to exceed the control limit by larger amounts, i.e., the monitoring performance of the B-P-type statistic improves with larger $w$.

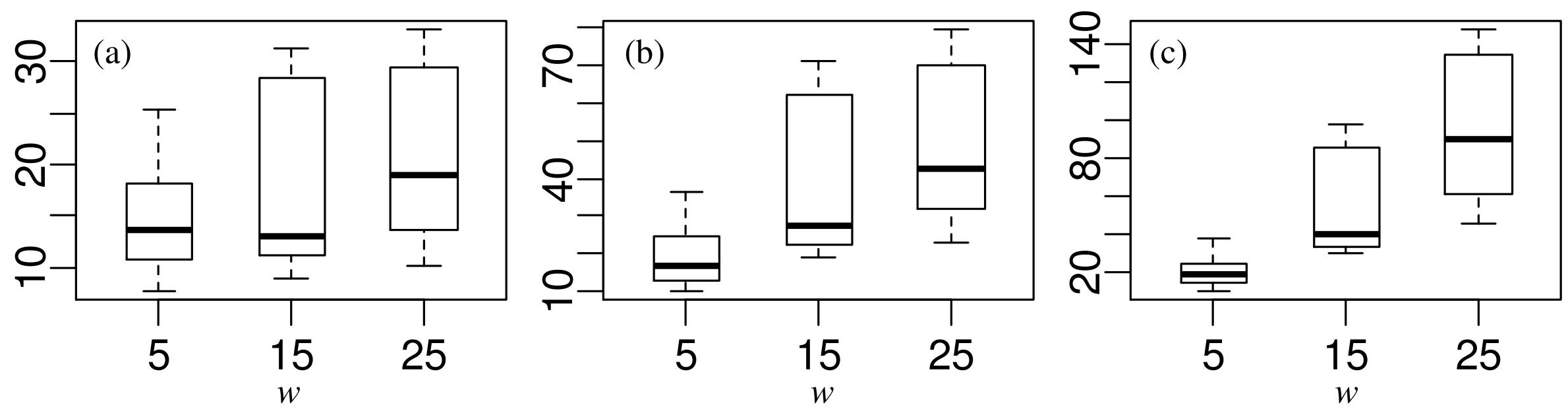

**Figure 7**. Boxplots of $(\bar{S} - CL)/(UCL - CL)$ across 10 replicates, where $\bar{S}$ is the average B-P-type monitoring statistic for all Phase II images containing defects of sizes: (a) 5×21, (b) 9×21, and (c) 15×21. Three window sizes $w$ = 5, 15, and 25 were considered.

However, using a larger $w$ for the B-P-type statistic has two potential drawbacks. First, a larger $w$ requires more computational expense, because the number of covariance terms to be computed in each moving window increases quadratically in $w$, and the kernel window is also larger. Second, and perhaps more seriously, using larger $w$ means that more boundary pixels (≈ $w/2$ pixels at each image edge) cannot be monitored, because full windows are required for computing the SMS. This is the reason why the monitoring performance of the B-P-based approach in Table 1 mildly degrades as $w$ increases for the smallest defects, which are more likely to occur at the boundary as $w$ increases. Therefore, for the B-P-based approach, we recommend that users choose $w$ as large as possible while balancing the above drawbacks.

We also compared our algorithm with three alternative methods. The first uses the Epanechnikov quadratic kernel in (6) to compute the weighted average of pixel intensities within a moving window, and this is used as the SMS. Specifically, the SMS of the $i^{th}$ pixel in an image is:

$$SMS_i = \frac{\sum_{h=-\infty}^{\infty}\sum_{m=-\infty}^{\infty} K(h,m) y_{i_1-h,i_2-m}}{\sum_{h=-\infty}^{\infty}\sum_{m=-\infty}^{\infty} K(h,m)},$$

where, as in Section 3.3, $i_1$ and $i_2$ are the row and column indices of the $i^{\text{th}}$ pixel. Similarly to our approach, the monitoring statistic for each image for this approach is the maximum $SMS_i$ (3) over all pixels in the image. We refer to this as the EPWMA approach. The second method, which we refer to as the EPWMV approach, is the same except that the SMS statistic is.

$$SMS_i = \frac{\sum_{h=-\infty}^{\infty}\sum_{m=-\infty}^{\infty} K(h,m) \left(y_{i_1-h,i_2-m} - \bar{y}_i\right)^2}{\sum_{h=-\infty}^{\infty}\sum_{m=-\infty}^{\infty} K(h,m)},$$

where $\bar{y}_i$ is unweighted mean of all pixel intensities in the window centered at pixel $i$.

The third method is the Haar-wavelet-based algorithm of Lin (2007a). Their algorithm divides a given image into many subimages and computes a monitoring statistic for each subimage, based on the 2-D Haar wavelet transform applied to these subimages. This is an example of the predefined-feature-based algorithms, where the features are defined by the 2-D Haar wavelet characteristics obtained from the subimages. The Lin (2007a) algorithm is not a standard control charting algorithm as defined in Megahed et al. (2011), because Lin (2007a) applies a spatial control chart within each image, as opposed to having a single charting statistic associated with each image. Thus, to have a common basis for comparison, we modify the Lin (2007a) approach as follows. The charted statistic for each image is taken to be the maximum of all of the Lin (2007a) statistics computed for all subimages of the image. Analogous to Table 1, Table 2 reports the average power at a Type I error rate of $\alpha = 0.003$ (the same $\alpha$ used for our approach) across 10 replicates, but for the EPWMA and EPWMV approaches (with $w$ = 5, 15, and 25) and the modified version of Lin (2007a). For each method, the control limits are chosen to control the Type I error based on a set of in-control images, analogous to how the control limit is selected for our algorithm. None of these approaches successfully detects the defects in this example.

**Table 2**. Average power across 10 replicates for the EPWMA, EPWMV, and Lin (2007a) methods for the same example depicted in Table 1.

| Defect sizes | EPWMA | | | EPWMV | | | Lin (2007a) |
|---|---|---|---|---|---|---|---|
| | $w = 5$ | $w = 15$ | $w = 25$ | $w = 5$ | $w = 15$ | $w = 25$ | |
| 5×5 | 0.005 | 0.004 | 0.006 | 0.137 | 0.006 | 0.007 | 0.005 |
| 5×21 | 0.004 | 0.005 | 0.002 | 0.123 | 0.005 | 0.003 | 0.006 |
| 9×21 | 0.007 | 0.004 | 0.003 | 0.110 | 0.002 | 0.004 | 0.007 |
| 15×21 | 0.005 | 0.005 | 0.006 | 0.078 | 0.008 | 0.002 | 0.012 |

### 5.2 Diagnostic stage

If the monitoring stage signals an alarm, the algorithm invokes the diagnostic stage, which uses the SMS values computed from Phase II of the monitoring stage and compares them with the diagnostic threshold(s) as discussed in Section 4.2. For illustration, in the second and third rows of Figure 6, we plotted the diagnostic images for the four defect-containing images in the top row of Figure 6, using A-D-based and B-P-type statistics, respectively, both with $w = 5$ (see the online supplement for this paper for analogous results for $w = 15$ and 25). To set the diagnostic threshold, we used $n_D = 10$ (equivalent to having an average of 10 noise-related black pixels in the in-control diagnostic images) and the empirical distribution of the SMS statistics computed for all pixels in all 1000 Phase I images.

## 6. Textile Application

Next, we apply our approach to a set of real image data for a textile material, an example image of which is shown in Figure 1(a). Note that the fabric pattern is quite complex (as a stochastic process) with random thicknesses of and distances between fiber strands. Figure 8 displays six images containing defects that were created by physically scuffing, tearing, or otherwise deforming the fibers locally to represent a variety of defect types. All images were also standardized as a preprocessing step in this example. All the image data in this example are available in the online supplement.

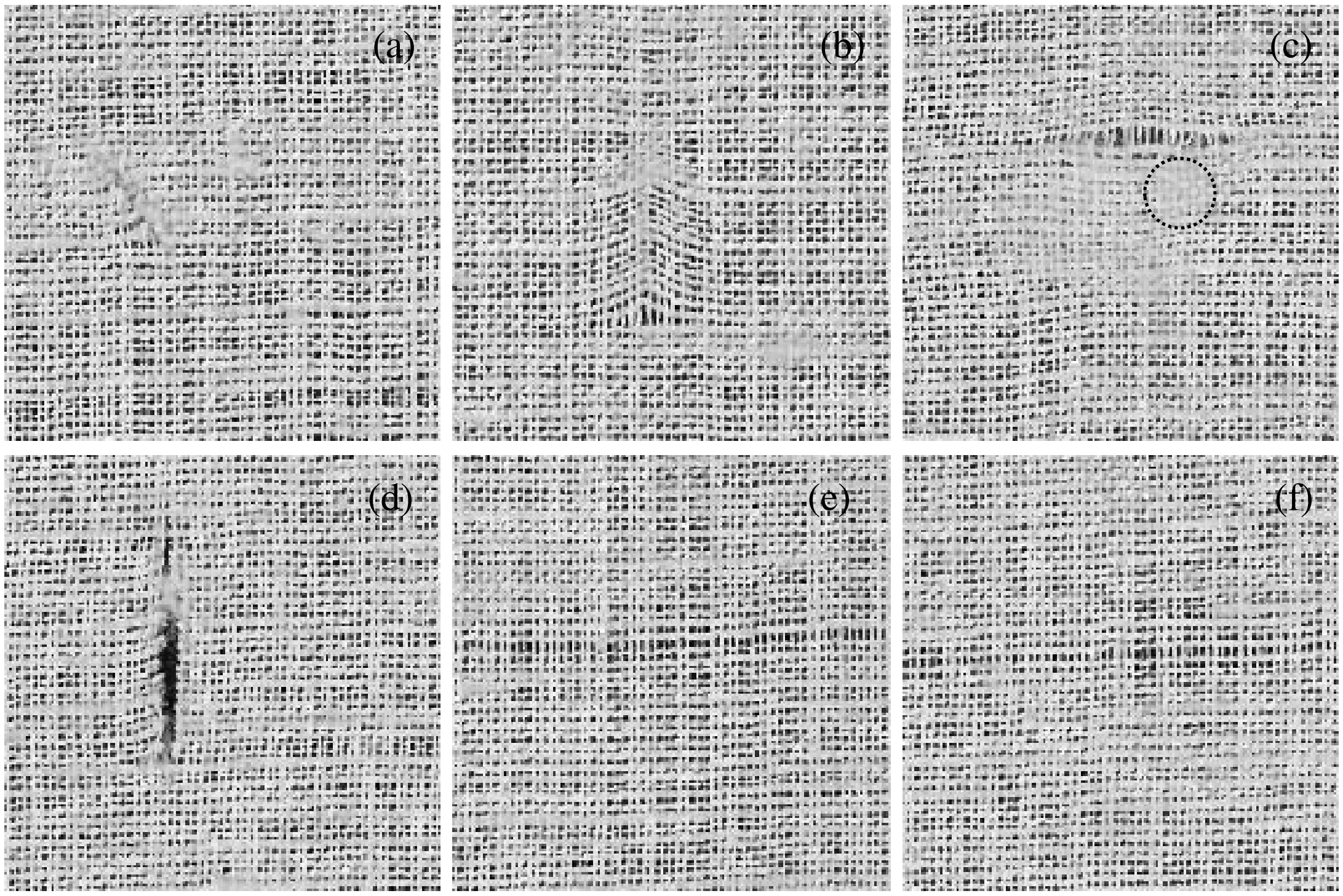

**Figure 8**. Defect-containing textile images corresponding to the in-control image in Figure 1(a), but with defects: (a) scratch, (b) fiber direction change, (c) tear, (d) hole, and (e—f) runs. The circled blurry region in panel (c) is an artifact of our imaging system and not a created defect, but it is more severe than the typical imaging blurs.

Regarding the validity of the MRF assumptions for this data set, the locality or "Markov" part virtually always holds if we select a large enough neighborhood size $l$. We use cross-validation within the supervised learning model fitting procedure to determine how large the neighborhood should be. The value of $l$ that minimizes the cross-validation error corresponds to the neighborhood size that includes all neighboring pixels that serve as useful predictor variables. In this respect, cross-validation identifies the neighborhood size that is required to make the stochastic surface Markov, which follows trivially by definition of the Markov property. In addition, the in-control images in this example passed the stationarity test for textured images of Taylor, Eckley, and Nunes (2014).

**6.1 Monitoring stage**

We used a regression tree as the supervised learning model in this example, and we used the image of size 500×500 pixels shown in Figure 1(a) to fit the regression tree. The neighborhood size $l$ of 15 was chosen by cross-validation. From the fitted regression tree, we computed the SMS values and the monitoring statistic $S$ in (3) for $N$ = 94 Phase I images, for both the A-D-based and B-P-based SMSs, using $w$ = 5, 15, and 25. For the A-D-based statistic, we chose $q_l \approx q_u \approx 1.8 \times 10^{-3}$ (corresponding to 400 observations in each tail), and $p \approx 2.2 \times 10^{-5}$ (corresponding to $r_p = -3.03$ and $r_{1-p} = 3.16$).

For this example, we also compare our algorithm with the EPWMA and EPWMV approaches and the modified version of Lin (2007a) described in Section 5. For all methods, we set the control limits based on the empirical distribution of their monitoring statistics, computed from the set of $N$ = 94 Phase I images, such that one of the 94 image statistics fell outside the control limits. For the EPWMA method, the control limits (LCL, UCL) were taken to be symmetric about the center line (CL), and likewise for the square root of the EPMVA statistic.

Table 3 reports the values of the CLs, LCLs, and UCLs and the monitoring statistics computed for the 6 defect-containing images in Figure 8 (which represent Phase II images) for all methods. The statistics that are beyond the control limits are in bold font. As was the case for the simulation example in Table 1, the monitoring performance of our algorithm when using the B-P-based statistic is slightly better than that when using the A-D-based statistic in this example. Nevertheless, with either monitoring statistic, the monitoring performance of our algorithm is clearly better than the performance of the other algorithms. As can be seen in Table 3, the EPWMA approach does not sound an alarm for any of the six defect-containing images in Figure 8. The Lin (2007a) algorithm only sounds an alarm for the most extreme defect in the image in Figure 8(d), whereas the EPWMV approach sounds an additional alarm for the image in Figure 8(f). In contrast, for most window sizes, our algorithm sounds an alarm for all six defect-containing images, and the signals often exceed the control limit by a large margin.

**Table 3**. Comparison of monitoring results in the textile example for our approach, and the EPWMA, EPWMV, and Lin (2007a) approaches for $\alpha = 1/94$. The last six columns shows the monitoring statistics computed for the six defect-containing images in Figure 8. Bold numbers indicate alarmed cases.

| Methods | | LCL | CL | UCL | (a) | (b) | (c) | (d) | (e) | (f) |
|---|---|---|---|---|---|---|---|---|---|---|
| A-D | $w = 5$ | $-\infty$ | 16.0 | 26.0 | **36.7** | **37.1** | **49.1** | **167** | **28.1** | **26.1** |
| | $w = 15$ | $-\infty$ | 18.8 | 29.1 | **37.0** | **56.1** | **56.2** | **453** | **34.5** | **34.4** |
| | $w = 25$ | $-\infty$ | 23.3 | 40.3 | **45.0** | **103.1** | **70.1** | **376** | 36.5 | 40.2 |
| B-P | $w = 5$ | $-\infty$ | 9.38 | 21.2 | 19.7 | **64.0** | **51.9** | **1346** | **27.4** | **29.7** |
| | $w = 15$ | $-\infty$ | 1.46 | 2.7 | **4.72** | **33.2** | **12.2** | **1730** | **5.87** | **7.17** |
| | $w = 25$ | $-\infty$ | 1.03 | 1.8 | **2.52** | **25.7** | **6.63** | **935** | **3.95** | **4.25** |
| EPWMA | $w = 5$ | 0.29 | 0.33 | 0.37 | 0.33 | 0.34 | 0.32 | 0.31 | 0.34 | 0.33 |
| | $w = 15$ | 0.06 | 0.12 | 0.18 | 0.13 | 0.11 | 0.13 | 0.10 | 0.10 | 0.10 |
| | $w = 25$ | 0.04 | 0.08 | 0.12 | 0.09 | 0.08 | 0.11 | 0.07 | 0.07 | 0.07 |
| EPWMV | $w = 5$ | 2.93 | 3.52 | 4.17 | 3.62 | 3.51 | **4.19** | 3.38 | 3.20 | 3.77 |
| | $w = 15$ | 1.72 | 2.08 | 2.48 | 2.22 | 2.28 | 2.27 | 2.43 | 2.29 | 2.40 |
| | $w = 25$ | 1.44 | 1.77 | 2.13 | 1.89 | 1.78 | 1.92 | **2.60** | 1.89 | **2.15** |
| Lin (2007a) | | $-\infty$ | 30.9 | 43.8 | 28.3 | 28.2 | 29.7 | **60.7** | 32.9 | 38.3 |

### 6.2 Diagnostic stage

In the following, we discuss the diagnostic results of our approach for the example in Section 6.1. As in the simulation example in Section 5, we used $n_D = 10$ to set the diagnostic threshold for all cases. The diagnostic images for the defect-containing images in Figures 8(a—e) are shown in Figure 9. The defect and diagnostic results for Figure 8(f) are similar to those for Figure 8(e), so we omit it here. The first and second rows of Figure 9 show the diagnostic images of our algorithm using the A-D-based statistic with $w$ = 5 and 25, respectively. Similarly, the third and fourth rows of Figure 9 show the diagnostic images of our algorithm using the B-P-type statistic with $w$ = 5 and 25, respectively. The results with $w$ = 15 for both statistics, which are somewhat in between the results with $w$ = 5 and $w$ = 25, can be found in the online supplement for this paper. The diagnostic images for the EPWMV approach with $w$ = 25 (the $w$ value that provided the best monitoring performance for EPWMV) and for the algorithm in Lin (2007a), constructed in the same manner as ours (using the same diagnostic threshold method described in Section 4.2), are also in the online supplement. In each row of Figure 9, the first—last columns are diagnostic images for Figures 8(a—e), respectively.

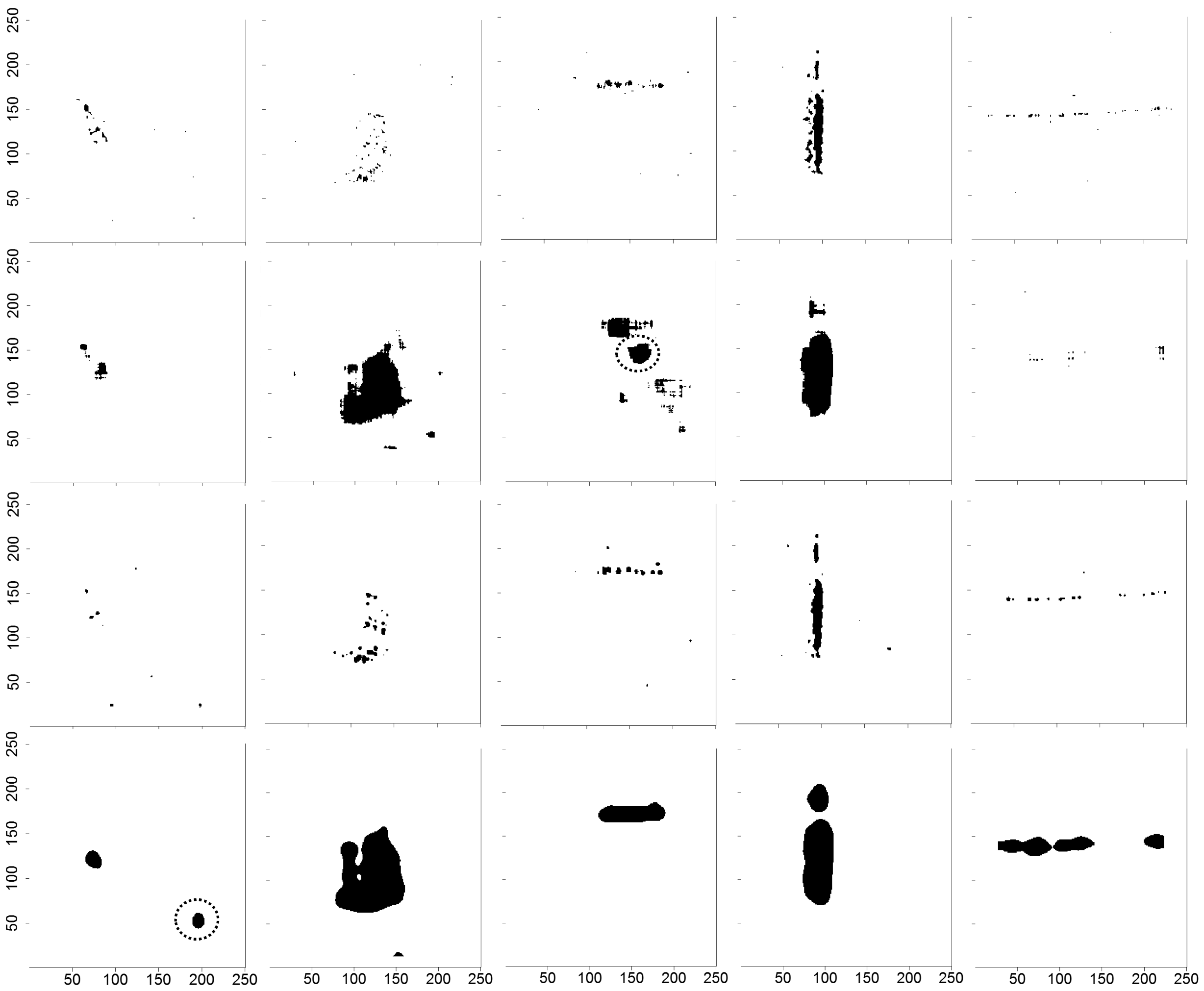

**Figure 9**. Diagnostic images of our algorithm for the defect-containing images in Figure 8 using: (1st row) A-D-based statistic with $w = 5$, (2nd row) A-D-based statistic with $w = 25$, (3rd row) B-P-type statistic with $w = 5$, and (4th row) B-P-type statistic with $w = 25$. In each row, the first—last columns correspond to Figures 8(a—e), respectively. The circled region is not a defect that we deliberately created, but it may indicate moderately abnormal local behavior in the fabric.

In general, our algorithm has correctly highlighted all the defects that we created, much more effectively than the EPWMV approach and the Lin (2007a) algorithm have. This is consistent with the comparison of the monitoring performances in Section 6.1. For our algorithm, both the A-D-based and B-P-based statistics worked quite well for diagnostic purposes, although B-P-based statistic may provide more pronounced highlights of the defects than the A-D-based statistic does.

There are some highlighted regions in the diagnostic images that do not correspond to the defects that we created, e.g., the circled regions in Figure 9. It is important to note that these are not false alarms in the conventional sense, because the purpose of the diagnostic stage is not to sound alarms. None of the highlighted regions that are not associated with the defects we created would have resulted in alarms in the monitoring stage for the control limits that were used for monitoring in Table 1, with the exception of the circled regions in Figure 9. Although we did not create these as defects, it appears that they are associated with moderately abnormal local behavior of the textile. The region circled in the diagnostic image in the bottom left corner of Figure 9 appears to have a relatively loose weave in Figure 8(a), and the circled region in the diagnostic image in the middle of the second row of Figure 9 is a blurry spot (also circled in Figure 8(c)) that is an artifact of our imaging system (which was not an industrial quality system) but that is larger and more pronounced than the typical blurs.

## 7. Conclusions

Stochastic textured surface data have a unique property that precludes the use of the existing SPC methods developed for profile data. Namely, existing profile SPC methods require a gold standard profile or at least a well-defined profile mean with meaningful features. On the other hand, most existing SPC methods applicable to stochastic textured surfaces seek to identify predefined features, which lack generality and are problem-specific by definition, requiring users' knowledge of the defects that are likely to occur. In contrast, we have developed a more general approach that is intended to detect any arbitrary local deviation from the normal in-control statistical behavior of the stochastic textured surfaces.

Our approach uses any appropriate off-the-shelf supervised learning algorithm to characterize the normal in-control statistical behavior of the stochastic textured surfaces. Based on the residuals of the fitted supervised learning model, we proposed two SMSs (A-D-based and B-P-based statistics) to quantify the local behavior of the residuals. We then use the max of the SMSs computed for all pixels in an image as the individual monitoring statistic for that image. We have illustrated the approach with examples of simulated stochastic textured surfaces and real textile fabric images. Both the A-D-based and the B-P-based statistics quite successfully detected and revealed (via the diagnostic images) the existence of defects of various natures. We have observed that the B-P-based statistic provided somewhat better performance than the A-D-based statistic in most of the examples.

There are a number of potential avenues to improve the performance of our approach. First, if the defects occur persistently across multiple images, the monitoring performance could likely be improved by accumulating our individual monitoring statistic $S_j$ using a EWMA or CUSUM type statistic, as mentioned in Section 3.1. Combining multiple charts, each with a different value of $w$, may also be useful to detect a wider range of defect sizes. Furthermore, for the B-P-based statistic, it may be useful to use a large value of $w$ over the interior part of the image (larger $w$ typically improves the monitoring performance of the B-P-based chart) and a smaller value of $w$ around the boundary of the images (because smaller $w$ allows the SMS's to be calculated closer to the boundary). Alternative choices of SMS and monitoring statistic, e. g., treating each neighborhood of residuals as a vector and then using some multivariate monitoring statistic on the residual vector as the SMS, could also potentially improve the performance. Likewise, more complex supervised learning models (e.g., boosted trees, random forests, deep neural networks, etc.) may also improve the performance, albeit at the cost of an increase in computational expense. In fact, we have tried boosted trees and neural networks, in addition to regression trees, but were able to achieve only moderate improvement (in terms of cross-validation error of the supervised learning model) in our examples. This may be because our computational limitations forced us to work with smaller size images and/or terminate the model fitting optimization algorithm early. Finally, the methods in Qiu and Yandell (1997) and Qiu (1998) might be useful for removing noise-related black pixels in the diagnostic images. We leave these for future studies.

## 8. Supplementary Materials

**Computer codes and data:** The proposed method in this paper has been implemented in the "spc4sts" package (Bui and Apley 2017a). The textile image data set used in this paper has a large size and has been included in a separate data package under the name "textile" (Bui and Apley 2017b). The tests of stationarity in the paper were conducted using the LS2Wstat package (Taylor and Nunes 2014).

**Further discussions and results**: We discuss the types of defects that our algorithm can detect. We also show diagnostic images for the simulation example when $w$ = 15 and 25 and for the textile example using our approach with $w$ = 15, the EPWMV approach with $w$ = 25 and the algorithm in Lin (2007a).

## Acknowledgement

This work was supported in part by NSF Grant # CMMI-1265709 and AFOSR Grant # FA9550-14-1-0032, which the authors gratefully acknowledge. Anh Tuan Bui also acknowledges support from the Vietnam Education Foundation. The authors thank the Editor and the anonymous Associate Editor and Referees for helping to improve the article.

Supplementary Materials

*A monitoring and diagnostic approach for stochastic textured surfaces*

Anh Tuan Bui and Daniel W. Apley

**What types of defects can be detected?**

This section aims to shed light on the types of the defects that our proposed algorithm can detect. Recall that our approach statistically characterizes the normal behavior of the stochastic textured surfaces via fitting a supervised learning model to a training image(s) that is representative of normal behavior, and we look for local changes in the behavior of the residuals of the fitted model. Consequently, the general condition required for defect detection by our algorithm is that the behavior of the pixels in the defect region are not consistent with what would be predicted based on their neighboring pixels; and that this causes a local change in the behavior of the residuals for the pixels in the defect region.

To illustrate, Figure S1(a) shows a simulated image for the same example used in Section 5, but we have added defects in the form of a black square and a white noise square. The former lie in the horizontal pixel indices 50—100, while the latter lie in the horizontal pixel indices 150—200. Both defects lie in the vertical pixel indices 101—151. Figure S1(b) plots the residuals along four lines that extend horizontally across the image, at vertical pixel indices 100, 101, 126, and 151. We label these four residual trace as trace 100, trace 101, trace 126, and trace 151. The solid circles and solid squares on the residual traces in Figure S1(b) correspond to the left/right boundaries of the black square and white noise square defects, respectively. The residual behavior clearly changes in the vicinity of the defects. For the black square defect, the residual mean is substantially different from normal in trace 100 and trace 151, whereas the residual variance is substantially smaller than normal in trace 101 and trace 126. For the white noise square defect, the residual variance is substantially larger than normal in all four traces. These deviations from the normal residual behavior in the vicinity of the defect region are reflected in the SMS statistics and thus are detected by our approach, as indicated in the diagnostic images in Figures S1(c) and (d).

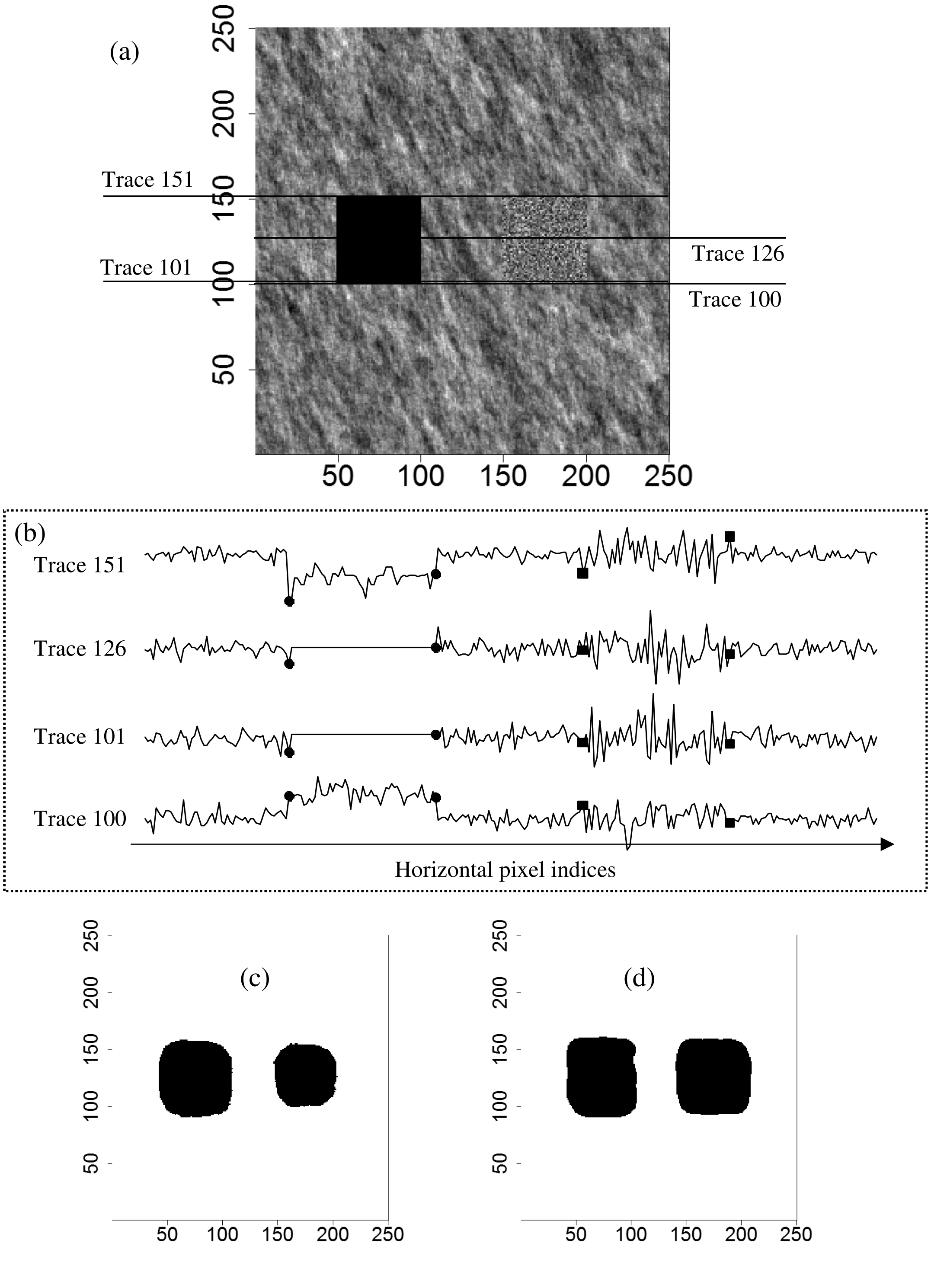

**Figure S1**. Behaviors of the residuals in the vicinity of defects: (a) an image containing a black square defect and a white noise square, (b) four residual traces across the four horizontal lines indicated in panel (a), corresponding to vertical pixel indices 100, 101, 126, and 151, (c) diagnostic image for panel (a) using the A-D-based statistic, and (d) diagnostic image for panel (a) using the B-P-based statistic. The residuals inside the pairs of solid circles and solid squares in panel (b) correspond to pixels within the black square and white noise square, respectively.

In theory, any deviation from the normal stochastic behavior of stochastic textured surfaces that results in a change in the residual behavior can be detected by our approach. In the spatial autoregressive process example of Section 5, our approach successfully detected the white noise defects, and it was also able to detect milder defects. For example, instead of reducing the $\phi_1$ and $\phi_2$ parameters of the spatial autoregressive process all the way to 0 (which corresponds to white noise), we can produce milder defects by only reducing these parameters by say 10% of their normal values of $\phi_1 = 0.6$ and $\phi_2 = 0.35$. Eighty milder-defect-containing images (20 images for each defect size) were generated in the same manner with the white noise defect except that $\phi_1 = 0.54$ and $\phi_2 = 0.315$. Figure S2 plots the monitoring statistic $S$ for these milder defect-containing Phase II images, as well as for some of the Phase I images, along with the control limit. The top three panels are for the A-D-based statistic for $w$ = 5, 15, and 25 (top to bottom) and the bottom three are for the B-P based statistic for the same three $w$. Again, our algorithm worked quite well for these mild defects with quite small sizes.

For the textile example, the results in Section 6 demonstrate that our approach is able to detect defects of various natures in this type of stochastic textured surface, including scratches in Figure 8(a), fiber direction changes in Figure 8(b), tears in Figure 8(c), holes in Figure 8(d), and runs in Figure 8(e) and Figure 8(f). It should be noted that our algorithm does not require that only a single type of defect is present in an image. The example in Figure S1 illustrates that it can detect the presence of multiple defects of different types.

**Additional diagnostic images for the simulation example**

Figure S3 plots the diagnostic images of the four defect-containing images that are shown in the first row of Figure 6. The first, second, third, and fourth rows of Figure S3 correspond to Figure 6(a), Figure 6(b), Figure 6(c), and Figure 6(d), respectively. The first, second, third, and fourth columns in each row of Figure S3 are results of A-D statistic with $w$ = 15, A-D statistic with $w$ = 25, B-P type statistic with $w$ = 15, and B-P type statistic with $w$ = 25, respectively.

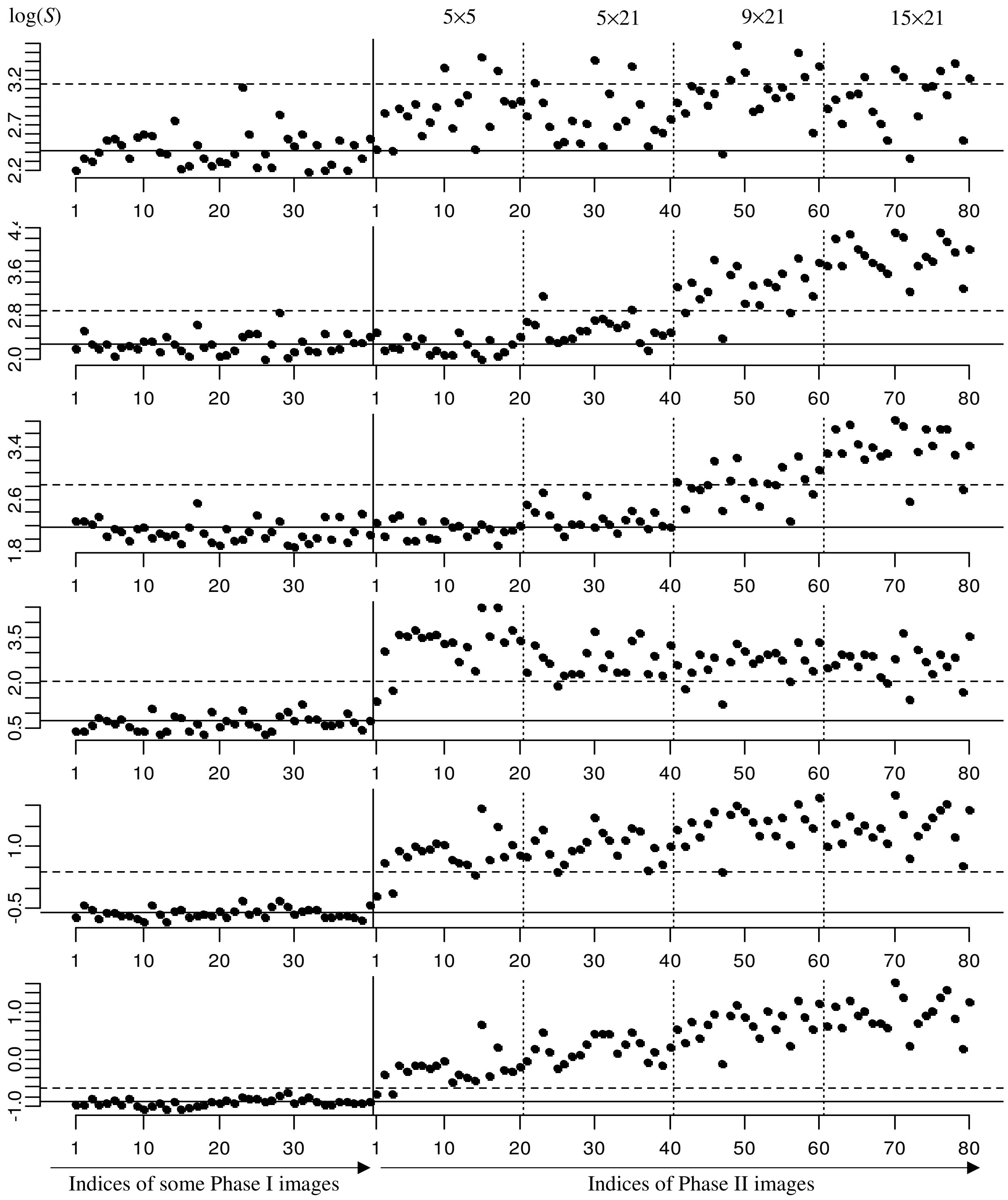

**Figure S2.** Monitoring results of our algorithm for the simulation example with milder defects. The top three panels and the bottom three panels correspond to A-D and B-P type statistics, respectively, using $w = 5$, 15, and 25, respectively (top-to-bottom). The vertical axes are the statistics $S$ in log scale. The vertical solid line at index $j = 1$ separates the Phase I and Phase II images, with the latter containing defects. The defect sizes in the first, second, third, and last set of 20 Phase II images are 5×5, 5×21, 9×21, and 15×21 pixels, respectively.

To set the diagnostic threshold, we use $n_D = 10$ (equivalent to having an average of 10 noise-related black pixels in the diagnostic images for in-control images) and the empirical distribution of the SMS statistics computed for all pixels in all 1000 Phase I images.

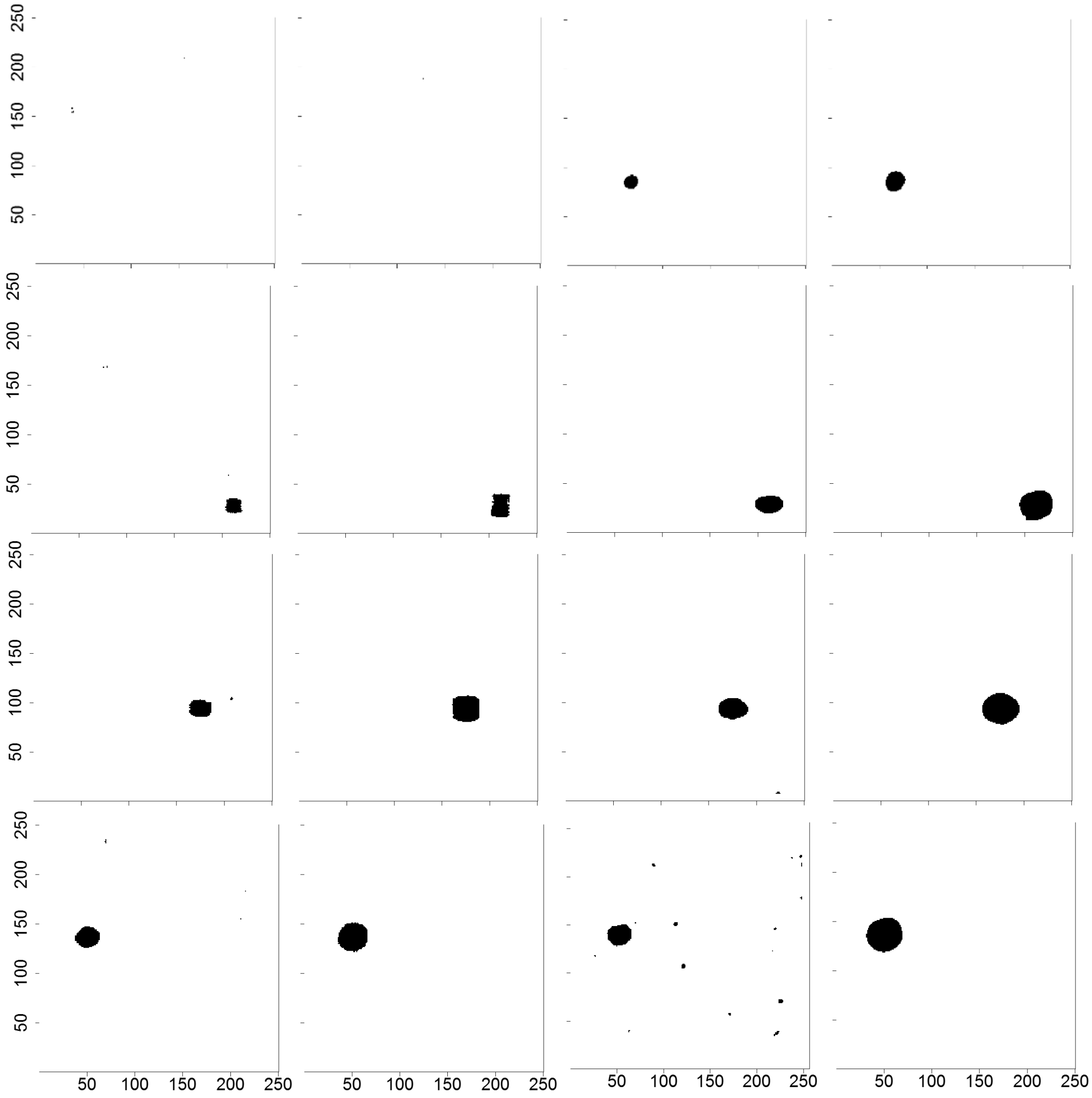

**Figure S3**. Diagnostic images of our algorithm for the defect-containing image in: (First row) Figure 6(a), (second row) Figure 6(b), (third row) Figure 6(c), and (fourth row) Figure 6(d). The results in each row are of: (first column) A-D statistic with *w* = 15, (second column) A-D statistic with *w* = 25, (third column) B-P type statistic with *w* = 15, and (fourth column) B-P type statistic with *w* = 25.

### Additional diagnostic images for the textile example

Figure S4 (first—last columns in each row) shows the diagnostic images for the defect-containing images in Figures 8(a—e). The first and second rows of Figure S4 show the diagnostic images of our algorithm using A-D-based and B-P-type statistics (both with $w$ = 15), respectively.

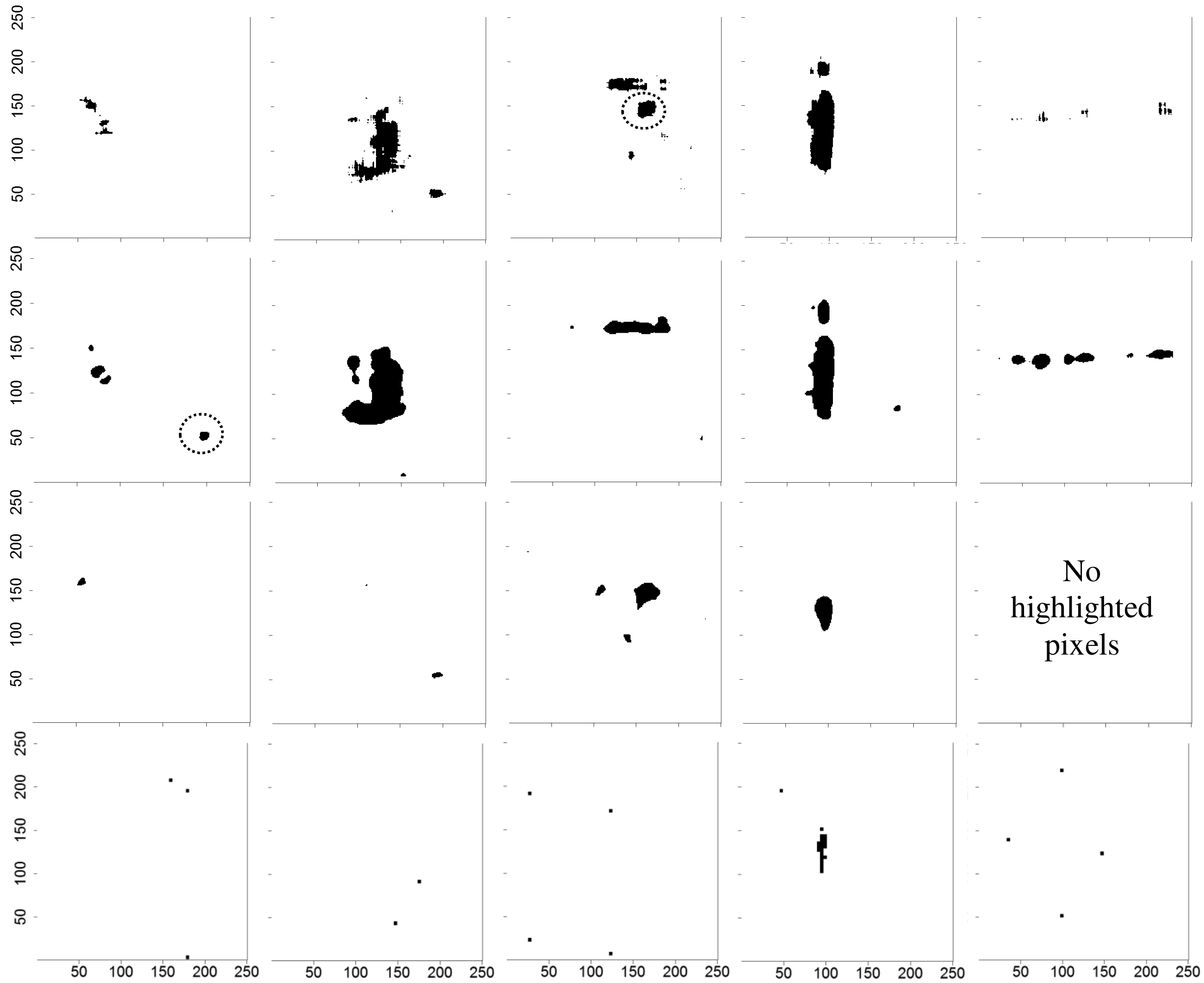

**Figure S4**. Diagnostic images for the defect-containing images in Figure 8 using: (first row) A-D-based statistic with $w$ = 15, (second row) B-P-type statistic with $w$ = 15, (third row) EPWMV algorithm with $w$ = 25, and (fourth row) Lin (2007a) algorithm. In each row, the first—last columns are diagnostic images for Figures 8(a—e), respectively. The circled region is not a defect that we deliberately created, but it may indicate moderately abnormal local behavior in the fabric.

Similarly, the third and fourth rows of Figure S4 show the diagnostic images of the EPWMV approach with $w$ = 25 and the algorithm in Lin (2007a), respectively. These diagnostic images are constructed in the same manner as ours, using $n_D = 10$ to set the diagnostic threshold.